\documentclass[12pt]{iopart}
\usepackage{comment}
\usepackage{CJKutf8}

\usepackage{graphicx}
\usepackage{amsfonts,amssymb}
\usepackage{caption}
\usepackage{subfig}
\usepackage{multirow}
\usepackage{hyperref}
\begin{document}

\title[Analysis of an adaptive lead weighted ResNet for multiclass classification of 12-lead ECGs]{Analysis of an adaptive lead weighted ResNet for multiclass classification of 12-lead ECGs}

\author{Z. Zhao$^{1,2}$, D. Murphy$^1$, H. Gifford$^3$, S. Williams$^4$, A. Darlington$^1$, S. Relton$^4$, H. Fang$^5$, D.C. Wong$^1$}

\address{$^1$ University of Manchester, UK $^2$ Xi'an Jiaotong University, Xi'an, China $^3$ University of Exeter, Exeter, UK $^4$ University of Leeds, Leeds, UK $^5$ Loughborough University, Loughborough, UK}
\ead{david.wong@manchester.ac.uk}
\vspace{10pt}
\begin{indented}
\item[]December 2021
\end{indented}

\begin{abstract}

\textbf{Background} - Twelve lead ECGs are a core diagnostic tool for cardiovascular diseases. Here, we describe and analyse an ensemble deep neural network architecture to classify 24 cardiac abnormalities from 12 lead ECGs.

\textbf{Method} - We proposed a squeeze and excite ResNet to automatically learn deep features from 12-lead ECGs, in order to identify 24 cardiac conditions. The deep features were augmented with age and gender features in the final fully connected layers. Output thresholds for each class were set using a constrained grid search. To determine why the model made incorrect predictions, two expert clinicians independently interpreted a random set of 100 misclassified ECGs concerning Left Axis Deviation.

\textbf{Results} - Using the bespoke weighted accuracy metric, we achieved a 5-fold cross-validation score of 0.684, and sensitivity and specificity of 0.758 and 0.969, respectively. We scored 0.520 on the full test data, and ranked 2nd out of 41 in the official challenge rankings. On a random set of misclassified ECGs, agreement between two clinicians and training labels was poor (clinician 1: $\kappa = -0.057$, clinician 2: $\kappa = -0.159$). In contrast, agreement between the clinicians was very high ($\kappa = 0.92$).

\textbf{Discussion} - The proposed prediction model performed well on the validation and hidden test data in comparison to models trained on the same data. We also discovered considerable inconsistency in training labels, which is likely to hinder development of more accurate models.
\end{abstract}

%
%
\submitto{Physiological Measurement}
%
%
%

\section{Introduction}
The 12-lead electrocardiogram (ECG) provides critical information that assists in identifying cardiac abnormalities. The signal from each of the 12 leads corresponds to the heart’s electrical activity from a distinct angle that can be mapped to the anatomy of the heart. A skilled interpreter can therefore use ECG signals from multiple leads to localise the source of a cardiac abnormality. Expert cardiologists can identify abnormalities with high accuracy. A recent systematic review highlighted how the accuracy of human expert interpretation may be as high as 95\% in a controlled setting in which the final diagnosis was known \cite{cook2020}.

However, expert-level human ECG interpretation is limited by the availability of a trained cardiologist and the time required to synthesize information from the 12-lead signal (and to document their findings). In clinical practice, the absence of cardiologists means that other, non-specialist, clinicians commonly make preliminary interpretations, but are demonstrably less accurate \cite{salerno2003}.

Computer-aided methods for ECG interpretation have been suggested as one approach for circumventing these resource constraints. Historically, the accuracy of these methods has been poorer than humans \cite{estes2013}. For instance, Anh et. al's 2006 reported how 19\% of atrial fibrillation were considered to be false positives when reviewed by a cardiologist \cite{anh2006}.

Traditional machine learning approaches, in which salient features of the ECG signal are first identified, have been successful for some use cases. As far back as 1991, the performance of some methods were almost as accurate as cardiologists, for a limited subset of clinical conditions \cite{willems1991}. However, such methods have frequently struggled to correctly interpret ECG with arrhythmias, conduction disorders and pacemaker rhythms \cite{schlapfer2017}.

Modern deep learning methods may be able to improve interpretation accuracy. Until recently, the use of such techniques for 12-lead ECGs has been impractical due to the shortage of labelled training data. There remains room for improvement over initial promising results \cite{ribeiro2020}. The public release of a new large labelled data set presents a fresh opportunity to revisit this challenging problem \cite{alday2020}.

Here, we consider the task of cardiac abnormality classification from 12-lead electrocardiogram (ECG) recordings of varying sampling frequency and duration. We have tackled this problem by developing a deep neural network architecture \cite{zhao2020}. Our architecture acknowledges the importance of the spatial relationship between the ECG channels by using a squeeze-and-excitation (SE) block. In this extended analysis, we present a deeper investigation into the strengths and weaknesses of this approach, including the use of expert clinical knowledge to determine why examples may be misclassified.

\section{Methods}
Our objective was to create a model that could accurately classify 12-lead ECG recordings into one or more of 27 clinical diagnoses. In practice, we considered only the 24 clinical diagnoses shown in Table 3 in the competition description paper \cite{alday2020}. Each class corresponds to a single ICD-10 code, with the exceptions of classes `PVC', `PAC', and `RBBB'. These classes corresponded to two clinical codes which we considered to be identical. The trained model and training code are available at: \url{https://github.com/ZhaoZhibin/Physionet2020model}.

\subsection{Dataset}
We trained our model used publicly-available data released for the 2020 Physionet challenge. Alday et al. provide detailed information about the data set \cite{alday2020}. In brief, the training data set contains 43,101 12-lead ECGs. ECG recordings are of variable duration (6 - 1800 s) and sampling frequency (257 - 1000 Hz), corresponding to variations in real-life practice. Each ECG was linked to a gender and age. Data were sourced from five hospitals; the number of recordings from each location varied between 72 and 10344 examples.

We tested the model on a hidden data set of 16630 ECGs. The set comprised of 6630 (40\%) ECGs collected from two of the same locations as the training set, and 10000 (60\%) from a sixth undisclosed location. The sixth location was an American institution, in which mean age, and ratio of male to female sex, were similar to the rest of the test set. However, the sampling frequency was lower (300 Hz) than the vast majority of the training data.

\subsection{Pre-processing}
We resampled all ECGs to the minimum frequency of 257 Hz using linear interpolation. To allow a fixed input size in the deep learning
model, each ECG was set to be 4096 points. During training, we ensured this by zero-padding any shorter duration signals and randomly
clipping any longer duration signals.

We scaled age into the range [0,1]. Both age and gender were encoded using one-hot encoding, with two additional mask variables to represent missing values (Fig. 2). No other pre-processing was undertaken. In particular, we highlight that the signals were not filtered, as we did not want to accidentally remove pertinent information. 

\subsection{Model architecture}
After obtaining the input signals, we designed an improved ResNet to assign the 12-lead ECG recordings into the 24 diagnostic classes.

\begin{figure}[htbp]
	\centering
	\includegraphics[width=15cm]{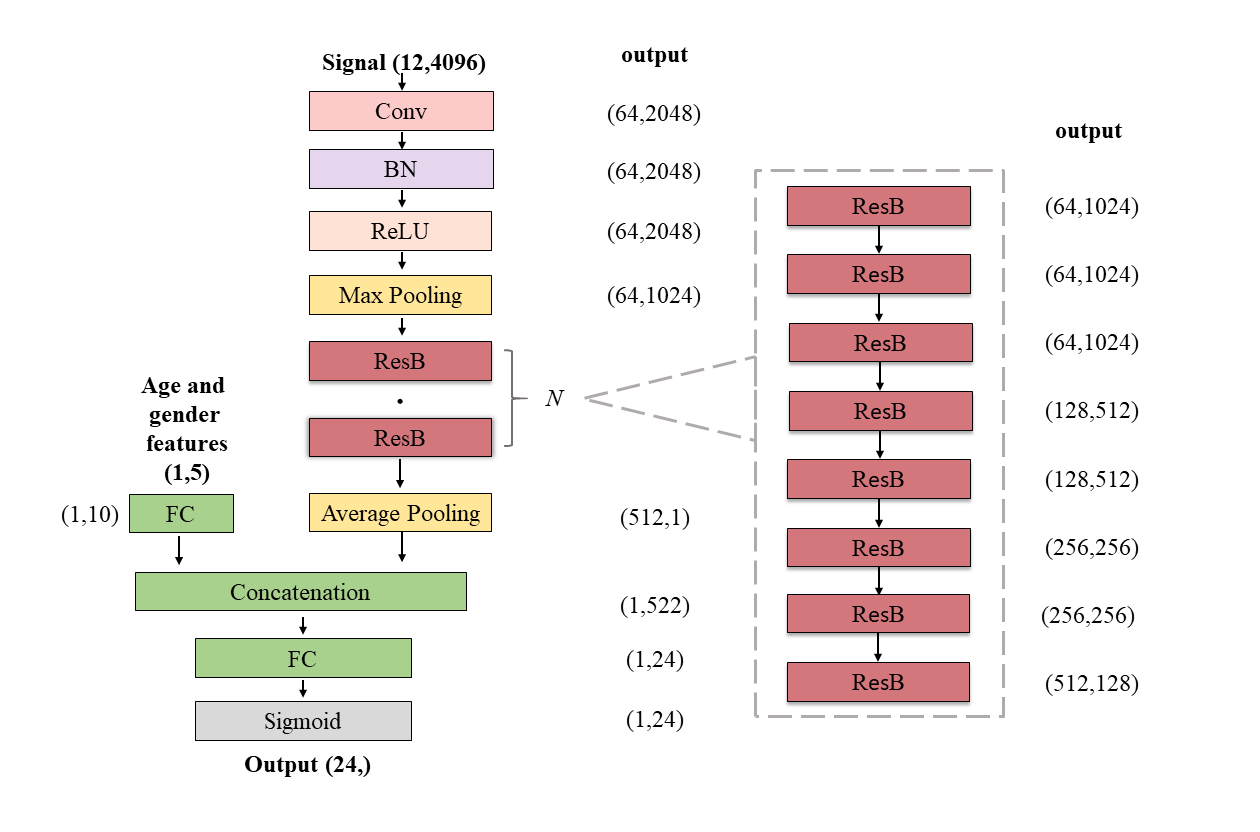}
	\caption{The proposed network architecture.}
	\label{model}
\end{figure}

The improved ResNet can be decomposed into \textit{feature extraction}, \textit{feature fusion}, and the \textit{classifier}. \textit{Feature extraction} consists of one convolutional layer followed by a batch normalization (BN) layer , a ReLU activation function, a max pooling layer, $N=8$ residual blocks (ResBs), each of which contains two convolutional layers and an SE block and an average pooling layer (Fig. \ref{model}). \textit{Feature fusion} concatenates deep features from the feature extraction part and the additional age and gender information. These combined features are input to the \textit{classifier}, which constitutes a Fully Connected (FC) layer and a Sigmoid layer, and outputs the probability of belonging to a disease class. An overview of our model is illustrated in Fig. \ref{model}.

Further details of the \textit{feature extraction} portion are as follows. The first layer of the network and the initial two ResB units have 64 convolution filters. The number of filters increases by a factor of two for every second ResB unit. The feature dimension is halved after the max pooling layer, and then after the fourth, sixth, and eighth ResBs.

We used a relatively large kernel size of 15 in the first convolution layer, and a kernel size of 7 in subsequent layers. Previous work has shown, empirically, that large convolutional kernels lead to better performance for ECG classification \cite{widekernel}.

\begin{figure}[htbp]
	\centering
	\includegraphics[width=10cm]{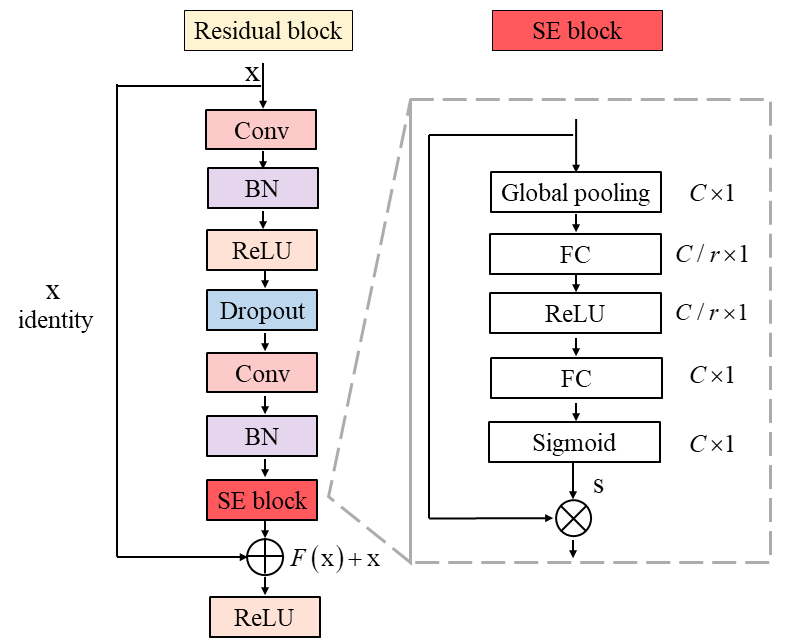}
	\caption{Residual block (ResB) and SE block.}
	\label{ResB}
\end{figure}

We chose stacked ResBs to extract features from ECG data as they are easy to optimize and have previously been effective in generating discriminative features \cite{he2016}. The structure of the modified ResB we use is illustrated in Fig. \ref{ResB}. It addresses the optimization problem by introducing a \emph{deep residual learning} framework. Instead of directly learning the underlying mapping $H(\mathbf{x})$, stacked layers in ResB approximate a residual function $F(\mathbf{x})=H(\mathbf{x})-\mathbf{x}$, where $\mathbf{x}$ is the input for ResB. Then the original function becomes $H(\mathbf{x})=F(\mathbf{x})+\mathbf{x}$.
This is easier to optimize as the identity mappings $H(\mathbf{x})=\mathbf{x}$ provides reasonable preconditioning.

We added a BN layer after each convolution layer to accelerate training. To reduce the likelihood of overfitting, we added a dropout layer with a drop out rate of 0.2 between the two convolutional kernels in each ResB.  

In developing our model architecture, we considered that, for many cardiac conditions, human interpretation of 12-lead ECG involves reviewing a subset of leads.
For instance expert clinical interpretation of Left Axis Deviation (LAD) is typically dependent on information from only three leads \cite{Hampton1997}.
We therefore chose an architecture that provides flexibility to model the relative importance of each lead.
One such approach, the SE block developed by Hu et al. has been previously applied to image data \cite{hu2018}.
Depicted in Fig. \ref{ResB}, we integrated the SE block to our ResB here for modeling the spatial relationship between the ECG channels.
First, global spatial information is \emph{squeezed} into a channel descriptor $\mathbf{z}\in\mathbb{R}^{C}$ by global average pooling, where $C$ is the channel dimension.
To make use of the information in $\mathbf{z}$, a simple gating mechanism with a Sigmoid activation is used to obtain the rescaling parameter $\mathbf{s}\in\mathbb{R}^{C}$:

\begin{equation}
\mathbf{s}
=
\sigma(g(\mathbf{z},\mathbf{W}))
=
\sigma(\mathbf{W}_2\delta(\mathbf{W}_1 \mathbf{z}))),
\end{equation}
where $\delta$ is the ReLU function, $\mathbf{W}_1\in\mathbb{R}^{\frac{C} {r} \times C}$, and $\mathbf{W}_2\in\mathbb{R}^{C\times \frac{C}{r} }$.
The parameter $r=16$ denotes the reduction factor, which controls the capacity and computational cost of the SE block.
The final output $\widetilde{\mathbf{x}} =[\mathbf{\widetilde{x}_1},\mathbf{\widetilde{x}_2},\dots,\mathbf{\widetilde{x}_C}]$ is calculated by channel-wise multiplication between channel features $\mathbf{x_C}$ and the scalar $s_c$:

\begin{equation}
\widetilde{\mathbf{x}}_C=\mathbf{x}_c\cdot s_c.
\end{equation}

The output of \textit{feature extraction} results in 512 deep features. These are concatenated with 10 age and gender features in an FC layer, which includes one hot encoding for gender and mask variables to denote missing data. We believed, a priori, that these were important features, given the clinical literature highlighting differing prevalence of certain heart conditions by age and gender (see, for instance, \cite{feinberg1995prevalence}) 

A final FC layer is used to complete classification from the 522 fused features. Finally, a Sigmoid function is used to scale classification results into [0, 1].

The training error for this multi-task problem was measured by the average binary cross-entropy loss.
The loss was optimized using the Adam optimizer with an initial learning rate 0.003.
The learning rate was reduced tenfold in the 20th and 40th epochs, and the model was trained for in total 50 epochs with a batch size of 64.

\subsection{Decision threshold optimisation}
The final layer of the proposed model is a $(24, 1)$ Sigmoid layer that estimates the probability that the signal belongs to a disease class.

Heavy class imbalance means that a default threshold of $P=0.5$ for each Sigmoid may be too insensitive. Rather than adjusting the model directly via the cost function, or by data resampling, we chose the pragmatic approach of adjusting the decision threshold. This assumes that the underlying representation, that is, the decision surface, is accurate. Kang et al. have previously argued that accurate representation is possible in problems, like ours, with class imbalance \cite{kang2019decoupling}.

For our multi-task problem, we determined thresholds by performing a constrained grid-search for each class individually. In detail, we first constrained thresholds for all classes to be equal. We set a temporary global threshold by searching in $[0,1]$ in steps of $0.1$. We then adjusted this threshold individually for each class, by searching in steps of $0.01$, with all other thresholds fixed. 

The drawback of this approach is that it makes the simplifying assumption that each class is independent. However, we decided that searching for the joint set of optimal thresholds would be too consuming to determine in this case, as it would require searching in 24-dimensional space.

\subsection{Model Analysis}
We used multi-label stratified five-fold cross-validation to assess the performance of the model.

\begin{figure}[htbp]
	\centering
	\includegraphics[width=10cm]{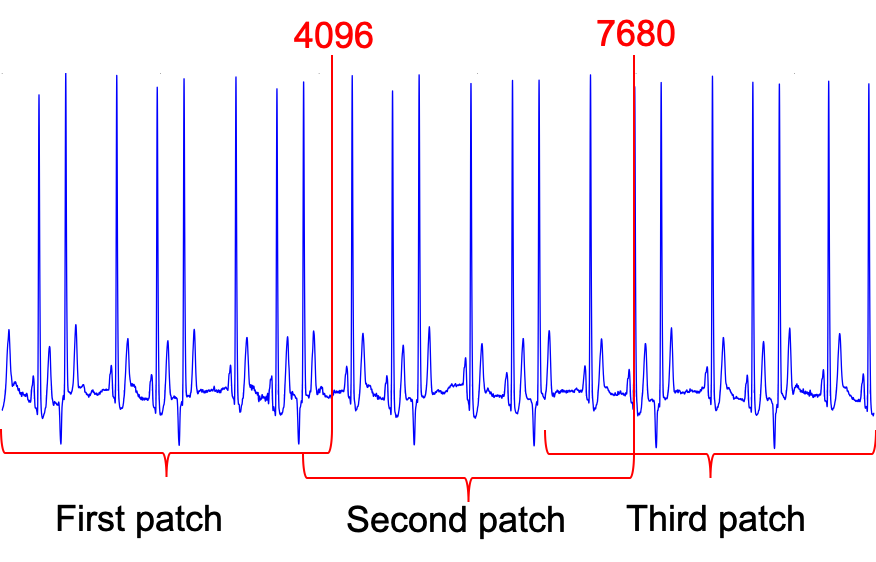}
	\caption{An example of segmenting the validation signals.}
	\label{val}
\end{figure}

For the validation and test signals, we continued to zero-pad any signals with fewer than 4096 samples. For signals longer than 4096 samples, we segmented the signals into multiple patches with a fixed overlap, $O = 256$. An example with sample length 10000 is depicted in Fig. \ref{val}. The number of patches, $P$, for a single signal can be formulated as:
\begin{equation}
P = ceil(\frac{L - 4096}{4096 - O}) + 1,
\end{equation}
where $ceil(\cdot)$ rounds a number upward to the nearest integer. We processed all $P$ patches, and used the mean of the output class probabilities to classify the recording.

In addition to sensitivity and specificity, we report model performance using a bespoke metric, $s_{normalized}$,
as described in \cite{alday2020}. This metric is a weighted accuracy that rewards incorrect classifications with similar risks or outcomes to the true class. 

Finally, we performed exploratory analysis to investigate the source of the classification errors. In post-hoc analysis, we selected a set of misclassified ECGs for a specific clinical condition, LAD. This condition was chosen as it is commonly determined using well-known heuristics, and we would therefore expect training labels for this condition to be reliable.

LAD occurs when the mean direction of the action potentials travelling through the ventricles at depolarisation (QRS axis) is less than $-30^{\circ}$. Archetypal examples are recognised via a positive QRS complex in Lead I, and negative QRS complexes in Leads II and III \cite{Hampton1997}.

A total of 100 examples (50 false positive and 50 false negatives) were selected at random. For each example, we asked two expert clinicians (HG,SW) with experience in ECG interpretation to determine whether LAD was present, not present, or whether it was unclear. We reported their agreement with each other and the training data labels using Cohen's $\kappa$.

\section{Results}
The final row of Table \ref{tab:result} states the $s_{normalized}$ metric for the intermediate validation set and hidden test set for our final model. In addition, we report 5-fold cross-validation estimates of sensitivity, specificity, f1-score and $s_{normalized}$ based on the training data. The first row reports the model metrics in which examples are assigned a class if P(class) $> 0.5$ (i.e. no threshold adjustment). The second row reports metrics for the final model with full threshold adjustment. Only this final model was scored on the hidden test set. The final model was submitted by the team ``\emph{between a ROC and a heart place}", and was 2nd of 41 models officially entered (and 2nd of 70 models submitted to the challenge).


\begin{table}[htbp]
\caption{Model results with different thresholds using five-fold cross-validation. model-default: an 
SE-ResNet without any threshold optimization; model-final: an improved ResNet with thresholds optimized by constrained grid-search.}
\vspace{4 mm}
\centerline{\begin{tabular}{l|ccc|c|c} \hline\hline
Method & \multicolumn{3}{c}{Training set} & Validation set & Test set\\
  & Sens. & Spec. & \(s_{normalized}\) & \(s_{normalized}\) & \(s_{normalized}\) \\ \hline
model-default  & 0.599 & 0.986	 & 0.630 &	0.607 &\\
\textbf{model-final}  &  \textbf{0.758} &	\textbf{0.969}	 & \textbf{0.684} &	\textbf{0.672} & \textbf{0.520}\\
\hline\hline
\end{tabular}}
\label{tab:result}
\end{table}

\subsection{Comparison to other models using validation set}

Per-class metrics for the validation set were provided by the challenge organisers. Table \ref{tab:font} summarises the average classification performance of the top 10 entries to the Physionet 2020 competition on the validation set, grouped by ECG class. All models had high f1-scores ($>0.8$) for Sinus Bradycardia (SB), Atrial Fibrillation (AF), Sinus Tachycardia (STach), and Complete Right Bundle Branch Block (CRBBB).

All models were poor ($f1 < 0.3$) at classifying Bradykinesia (Brady), Non-Specific Intra-Ventricular Conduction Delay (NSIVCD), Pacing Rhythm (PR), Pre-Ventricular Contraction (PVC), Right Axis Deviation (RAD), and T-wave Inversion (TInv). In most cases, poor performance can be explained by the relatively small number of training and validation examples. For instance, in the case of PR, there were 299 training and 2 validation examples.

The limited number of validation samples for Brady also explains the difference in performance between the classification of Brady ($f1 = 0.003, n = 1$) and  Sinus Bradycardia (SB) ($f1 = 0.91, n = 860$). Given that SB is a common sub-type of Brady, we might otherwise expect classification metrics to be similar.

For others conditions, such as NSIVCD, where manual assessment is known to be challenging \cite{eschalier2015}, it is unsurprising that automated methods have poor performance. This will occur if either the key clinical features are unable to be represented by a deep neural network (e.g. if the features are too nuanced), or if the wrong features are encoded (e.g. if training data mislabelled). The exception to this is TInv (f1 = $0.29, n = 438$). The uniform poor performance over all models is surprising, given that there many examples, and that the clinical feature is simple to recognise for human experts. 

In comparison to the other top 10 models, our model had similar predictive power for all classes - the f-score was within 1 s.d. of the mean, for every class. More detailed per-class comparison of the models is not helpful, as minor differences may be due to model overfitting. Indeed, the model that generalised best to the test data had the lowest per-class metrics in the validation data \cite{natarajan2020wide}.

We further note that our model architecture was very similar to the entry that placed third in the 2020 Physionet challenge \cite{zhu2021identification}. The primary difference between the two models, which may explain difference in performance, was our inclusion of age and gender features. 

\begin{table}[htbp]

\caption{mean and standard deviation f-measure of each class, alongside number of training and validation examples, for the 10 top performing ($s_{normalized}$) entries to the 2020 Physionet challenge. Emboldened items indicate conditions that were poorly classified by all entries.}

\vspace{4 mm}
\centerline{
\begin{tabular}{cccc} \hline\hline
Class (training n, validation n) & SE-net &  Top 10 Mean f-1 & Top 10 std f-1 \\ \hline
IAVB (2394, 552) & 0.736 & 0.756 &	0.027\\
AF (3475, 552)  & 0.801  & 0.810 &	0.022 \\
AFL (314, 109) & 0.467 & 0.473 &	0.081\\
\textbf{Brady (288, 1)}& \textbf{0.000} & \textbf{0.003} &	\textbf{0.008}\\
CRBBB (683, 18)& 0.817 & 0.815 &	0.027 \\
IRBBB (1611, 206)& 0.495 & 0.463 &	0.060\\
LAnFB (1806, 110)& 0.428 & 0.391 &	0.079\\
LAD   (6086, 478)& 0.672 & 0.636 &	0.053 \\
LBBB (1041, 156) & 0.752 & 0.754 &	0.031\\
LQRSV (556, 192) & 0.573 & 0.393 &	0.177\\
\textbf{NSIVCB (997, 96)} & \textbf{0.133} & \textbf{0.13} &	\textbf{0.057} \\
\textbf{PR (299, 2)} & \textbf{0.000} & \textbf{0.041} &	\textbf{0.061}\\
PAC (1729, 459) & 0.618 & 0.592 & 0.061\\
\textbf{PVC (188, 178)} & \textbf{0.301} & \textbf{0.286} &	\textbf{0.048} \\
LQT (1513, 740) & 0.585 & 0.570 &	0.059\\
QAb (1013, 239) &	 & 0.362 &	0.073\\
\textbf{RAD (427, 38)} &  \textbf{0.208} & \textbf{0.243} &	\textbf{0.056} \\
SA (1240, 236) & 0.624 & 0.598 &	0.089\\
SB (2359, 860) & 0.934 & 0.908 &	0.036\\
SNR (20846, 1100)& 0.670 & 0.651 &	0.044 \\
STach (2402, 648)& 0.844 & 0.857 &	0.022\\
TAb (4673, 1119)& 0.626 & 0.595 &	0.047\\
\textbf{TInv(1112, 438)} & \textbf{0.248} & \textbf{0.201} & \textbf{0.060} \\
\hline\hline
\end{tabular}
}
\label{tab:font}
\end{table}

\subsection{Model error examples}
Having established that our model performed relatively well overall, we then investigated individual examples that our model misclassified.

\begin{figure}
  \centering
  \begin{tabular}{@{}c@{}}
    \includegraphics[width=15cm]{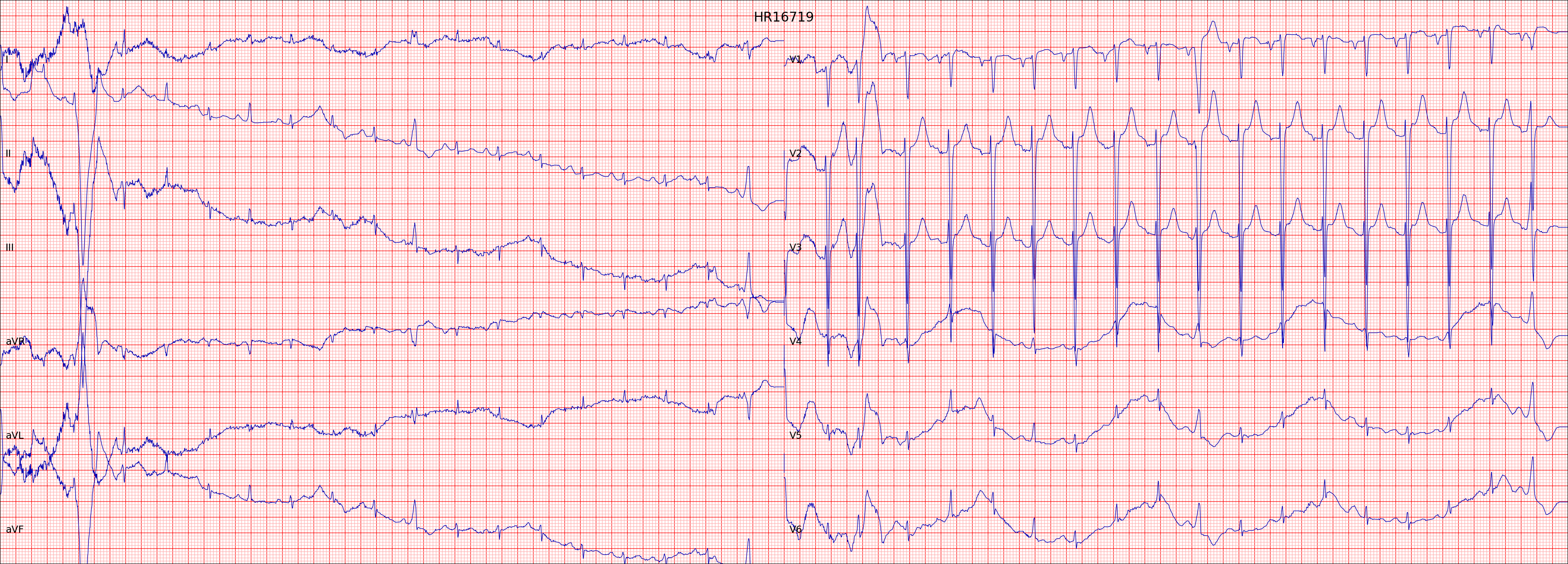} \\[\abovecaptionskip]
    \small (a) training label = LAD, model prediction = no LAD
  \end{tabular}

  \vspace{\floatsep}

  \begin{tabular}{@{}c@{}}
    \includegraphics[width=15cm]{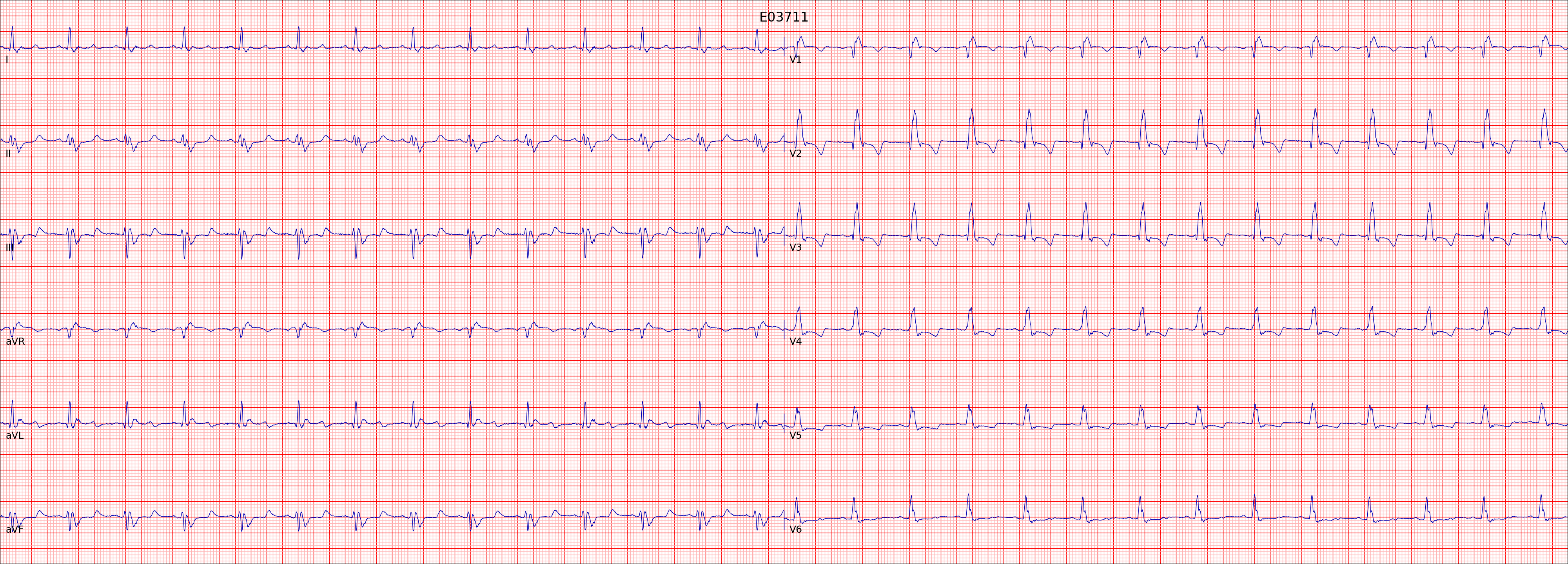} \\[\abovecaptionskip]
    \small (b) training label = no LAD, model prediction = LAD
  \end{tabular}

  \caption{Examples of ECGs misclassified by the SE-Resnet model.}\label{fig:ECGs}
\end{figure}

Figure \ref{fig:ECGs} (a) depicts an example that was labelled as having LAD in the training set, but was not identified by our model (i.e. a false negative). In this case, two independent clinical expert reviewers were both unable to determine whether LAD was present, due to low signal-to-noise ratio in the key I, II and III leads. Figure \ref{fig:ECGs} (b) depicts an example that was not labelled with LAD, but our model classified it is as LAD (a false positive). Additional clinical review (HG, SW) determined that this should have been labelled with LAD, due to the presence of a positive R peaks in Lead I and negative R peaks in Leads II and III.

From visual inspection of these, and other similar examples, we noted that discordance occurred because the neural network failed to recognise the key indicators of LAD (either due to noisy data, or inadequate representation) or that the neural network was correct, but the examples appear to be mislabelled. These two problems will place an upper-bound on the performance of any deep-learning model.

To explore this further and provide an initial attempt to quantify this problem, we randomly selected 50 examples of false-positive LAD and 50 examples of false-negative LAD. Two clinical experts classified the examples into \{\textit{LAD, no LAD, unsure}\}.

Table \ref{tab:clin_vs_clin} shows the interrater agreement between the two clinicians. Of the 100 examples, at least one clinician was \textit{unsure} in 18 cases. In the remaining 82 examples, the clinicians disagreed in only two instance - resulting in Cohen's $\kappa = 0.92$.

\begin{table}[htbp]
\caption{Comparison of clinican 1 (HG) vs clinician 2 (SW) for classification of 50 FP and 50 FN Left Axis Deviation (LAD) examples. $\overline{LAD}$ represents a classification of \textit{no} LAD.}

\vspace{4 mm}
\centerline{\begin{tabular}{ll|ccc} \hline\hline
 & & \multicolumn{3}{c}{Clinician 1 (HG)}\\
  & &LAD & $\overline{LAD}$ & Unsure \\ \hline
\multirow{3}{*}{Clinician 2 (SW)}& LAD  & 7 & 38 & 5\\
& $\overline{LAD}$  & 9 & 33 & 8\\
& Unsure  & 9 & 33 & 8\\
\hline\hline
\end{tabular}}
\label{tab:clin_vs_clin}
\end{table}

In contrast, both clinicians commonly disagreed with the training labels (Table \ref{tab:clin_vs_lab}. HG disagreed in 47/87 cases (Cohen's $\kappa = -0.057$) and SW disagreed in 53/91 (Cohen's $\kappa = -0.159$), excluding cases in which the clinicians were unsure. For both clinicians, disagreements mainly occurred when the example was labelled with LAD, but the clinicians, and model, did not believe there was LAD (i.e. False Negatives).

\begin{table}[htbp]
\caption{Comparison of clinican experts {HG,SW} classification with training set labels for Left Axis Deviation (LAD). $\overline{LAD}$ represents a classification or label of \textit{no} LAD.}
\vspace{4 mm}
\centerline{\begin{tabular}{ll|ccc|ccc} \hline\hline
 & & \multicolumn{3}{c}{Clinician 1 (HG)} & \multicolumn{3}{c}{Clinician 2 (SW)}\\
  & & LAD & $\overline{LAD}$ & Unsure & LAD & $\overline{LAD}$ & Unsure \\ \hline
\multirow{2}{*}{Model} & LAD  & 7 & 38 & 5 & 8 & 38 & 4\\
& $\overline{LAD}$  & 9 & 33 & 8 & 15 & 30 & 5 \\
\hline\hline
\end{tabular}}
\label{tab:clin_vs_lab}
\end{table}

\section{Discussion}
This paper proposes the use of an modified Resnet for multiclass ECG classification. The main modification is the addition of an SE layer, which allows better modelling of channel interdependencies.

The model performed well, placing 2nd out of 41 teams in the 2020 Physionet Challenge. However, like many other deep neural network architectures trained on the Physionet Challenge 2020 data, our model did not generalise well to the hidden test data. This may be due to overfitting, and we noted that 2020's winning entry incorporated hand-crafted features that would increase model robustness.

By analysing a small convenience sample of training data examples of one cardiac condition, LAD, we observed instances that appeared to be mislabelled. To estimate the extent of possible mislabelling, we extracted a larger sample of training data that have been misclassified by our model. Two clinicians independently assessed the sample. We found that the clinician's had very high agreement with each other ($\kappa = 0.92$), but very poor agreement with the training labels (HG: $\kappa = -0.057$, SW: $\kappa = -0.159$). The implication of this for machine learning models is that there is an upper-bound on their accuracy which may limit the possibility of creating a generalisable model that performs at human expert level in all settings.


The two problems of model overfitting and unreliable training data might both be caused by the same underlying issue. Even though the Physionet ECG data has been sourced from many locations, the differences in size of the datasets means that any resulting deep learning model will tend to be dominated by the largest data sets - leading to overfitting. Similarly, differences in the reliability of training labels may not be random, but dataset-specific. Indeed, for our example of LAD, we are aware that the presence of QRS-complex inversion in Lead II is a requirement for diagnosis of LAD in UK text books, but not in some Chinese texts [\cite{Hampton1997, diagnostics_2013}].

One approach to dealing with this problem may be create models that account for differences between datasets. More generally, this approach is known as domain adaptation. One specific way to implement this, initially proposed by Alvi et al \cite{alvi2018turning}, is so-called `Joint Learning and Unlearning'. This uses an adversarial multi-task approach to simultaneously minimise domain (i.e.~data set) prediction accuracy and maximise task accuracy. This approach has been successfully used for MRI segmentation problems, and we have begun investigating how it may be used for ECG data \cite{shang2021}.

In conclusion, we have trained an SE-net that automatically identifies 24 cardiac abnormalities from 12-lead ECG. In validation on external data, the model performed much worse than in internal cross-validation. Further analysis of the model outcomes suggests that this may be partly due to difference in labelling procedures between different training data sets.

\section*{References}
\bibliography{main.bib}
\bibliographystyle{unsrt}

\end{document}